\documentclass[prl,twocolumn,superscriptaddress,preprintnumbers,nofootinbib,floatfix]{revtex4-2}
\usepackage{epsfig}
\usepackage{graphicx,physymb}
\usepackage{hyphenat}
\usepackage{amsmath}
\usepackage{comment}
\usepackage{amssymb}
\usepackage{mathtools}
\usepackage{mathrsfs}
\usepackage{slashed}
\usepackage{epstopdf}
\usepackage{xcolor}
\usepackage{braket}
\usepackage{booktabs}
\definecolor{lcolor}{rgb}{0.,0.0,0.}
\definecolor{citcolor}{rgb}{0,0.,0.5}
\usepackage[breaklinks,colorlinks,urlcolor=blue,citecolor=blue,linkcolor=blue]{hyperref}
\usepackage{multirow}
\usepackage{ltablex}
\usepackage{soul}
\usepackage[utf8]{inputenc}
\usepackage[T1]{fontenc}
\usepackage{latexsym}
\usepackage[normalem]{ulem}

\def\CP{{\cal P}}

\newcommand{\beq}{\begin{eqnarray}}
\newcommand{\eeq}{\end{eqnarray}}

\newcommand{\bem}{\begin{multline}}
\newcommand{\eem}{\end{multline}}
\newcommand{\beg}{\begin{gather}}
\newcommand{\eeg}{\end{gather}}

\newcommand{\nn}{\nonumber\\}

\newcommand{\ben}{\begin{eqnarray*}}
\newcommand{\een}{\end{eqnarray*}}

\newcommand{\secn}[1]{Section~1}
\newcommand{\appn}[1]{Appendix~1}

\long\def\comment#1{ }

\def\and{\quad\text{and}\quad}

\def\0{{\boldsymbol 0}}

\def\0{{\boldsymbol 0}}

\begin{document}

\title{Exact group invariant scar towers in two dimensional  gauge theories}

\author{Jo\~{a}o Barata}
\email{joao.lourenco.henriques.barata@cern.ch}
\affiliation{CERN, Theoretical Physics Department, CH-1211, Geneva 23, Switzerland}

\author{Kiryl Pakrouski}
\email{pkiryl@ethz.ch}
\affiliation{Institute for Theoretical Physics, ETH Zurich, 8093 Zurich, Switzerland}

\author{Andrey V. Sadofyev}
\email{andrey.sadofyev@ehu.eus}
\affiliation{Department of Physics, University of the Basque Country UPV/EHU, P.O. Box 644, 48080 Bilbao, Spain}
\affiliation{IKERBASQUE, Basque Foundation for Science, Plaza Euskadi 5, 48009 Bilbao, Spain}

\preprint{CERN-TH-2026-209}

\begin{abstract}
We construct exact many-body scar towers in two dimensional gauge theories with two flavors of massless fundamental fermions. For oppositely charged fermions, the scar subspace is generated by a gauge-neutral $\eta$-pairing operator, and admits a purely algebraic construction. Exact diagonalization methods reveal anomalously low entanglement and long-range pair correlations in the scar states. 
Charge conjugation maps the $\eta$-tower to a vector-flavor polarization sector of the equal charge model, yielding an 
interpretation as a coherent flavor mode with Josephson-like phase dynamics. Finite fermion masses mix the protected pair with an orthogonal channel and destroy the exact tower. The identified scar tower can be algebraically realized in a large family of lattice field theories, revealing a universal character
of these states. Our results provide an analytical construction of invariant scar subspaces in a gauge theory with a nontrivial continuum interpretation.
\end{abstract}

\maketitle

\textit{Introduction:} Understanding how out-of-equilibrium states lose memory of their initial conditions and approach thermal equilibrium is a central problem in quantum gauge field theories, 
see e.g.~\cite{Berges:2020fwq} for a recent review. In generic interacting many-body systems, this process is commonly treated within the Eigenstate Thermalization Hypothesis (ETH) framework, according to which individual finite energy density eigenstates reproduce thermal expectation values of local observables~\cite{deutsch1991quantum,srednicki1994chaos,rigol2008thermalization,Deutsch:2018ulr,DAlessio:2015qtq}. 
Directly 
testing ETH in field theory is nontrivial~\cite{Delacretaz:2022ojg,Srdinsek:2023yli,CortesCubero:2019ann}, as is extrapolating 
lattice results to the thermodynamic and continuum limits, see e.g.~\cite{Delacretaz:2022ojg,Desaules:2022ibp,Desaules:2022kse}. Although signatures of thermalization have been 
observed in lattice and continuum gauge theories~\cite{Zhou:2021kdl,Wang:2022dpp,ALICE:2010suc,ALICE:2015xmh}, whether generic eigenstates are thermal, and under which conditions exceptional athermal states persist, remains an open question.

Quantum many-body scars provide a setting in which some of these questions can
be addressed. Scarred eigenstates are atypical, nonthermal eigenstates embedded
in an otherwise thermal spectrum. They are commonly distinguished by anomalously
low entanglement, nonthermal expectation values and correlations, and coherent
revivals from initial states having enhanced overlap with the scarred
subspace~\cite{Turner:2017fxc,Serbyn:2020wys,Moudgalya:2021xlu}. Signatures of many-body scarring have been observed in different experimental quantum
platforms~\cite{Bernien:2017ubn,Su:2022glk}, and a variety of mechanisms for generating
scar subspaces have been identified~\cite{Shiraishi:2017tsj,ODea:2020ooe,James2018,
Serbyn:2020wys,Moudgalya:2021xlu}, some \cite{MoudgalyaFragmCommutant22,Moudgalya:2022nll} also related to the adjacent field of Hilbert space fragmentation \cite{Khemani:2019vor,PhysRevX.9.021003,Sala_2020,budde2026hilbertspacefragmentationgeneralized,budde2026hilbertspacefragmentationgauge}. Nevertheless, analytically establishing the existence of such exceptional states is challenging. Moreover, the relevance and existence of scar states is commonly associated to finite size systems or truncated theories, while their description in the continuum is less clear, and seldomly discussed~\cite{Desaules:2022ibp,Desaules:2022kse,James2018,Delacretaz:2022ojg,Srdinsek:2023yli}.

In this \textit{Letter}, we show that exact quantum scar states can be constructed in two dimensional gauge theories, admitting a many-body interpretation. These exact scar states arise from spectrum-generating algebras, being associated to group invariant subspaces annihilated by part of the Hamiltonian~\cite{pakrouski2020GroupInvariantScars,Pakrouski:2021jon}. This construction is conceptually distinct from most of the literature on many-body scars in lattice gauge theories, see e.g.~\cite{Aramthottil:2022jvs,Desaules:2022ibp,
Desaules:2022kse,Halimeh:2022rwu,budde2024quantummanybodyscarsarbitrary,budde2026spectrumgeneratingalgebrahigherdimensional,Hartse:2024qrv,Calajo:2024bvs,Andrade:2024iqu}, typically relying on truncated representations of the continuum theory, with scar subspaces determined from spectral or dynamical diagnostics on the lattice.

Our discussion is motivated by, although not restricted to, the massless two flavor Schwinger
model~\cite{Schwinger:1962tp} with oppositely $q$-charged fundamental fermions, i.e. QED$_2^{(q,-q)}$~\cite{Coleman:1976uz,Schwinger:1962tp}. The scar tower is associated with the local gauge singlet pair operator 
\begin{equation}
    \eta^{\prime\dagger}
    = \sum_{j=1}^{L}(-1)^j
      \chi^\dagger_{+,j}\chi^\dagger_{-,j}\, ,
    \label{eq:eta_intro}
\end{equation}
with $\chi_{\alpha ,j}$ the single fermion field of the $\alpha$ charged field at position $j=1,2,\cdots , L$. When acting on the neutral zero-flux reference state $\lvert\Omega\rangle$, it
generates the number-projected condensates
\begin{equation}
    \lvert n^{\eta^\prime}\rangle
    = \frac{(\eta^{\prime\dagger})^n}{\sqrt{\mathcal N_n}}
      \lvert\Omega\rangle\, ,
    \label{eq:tower_intro}
\end{equation}
with $n=0, \, \cdots, L$, $    \mathcal N_n=(n!)^2\binom{L}{n}$, and corresponding to a scar tower in this model. We note that this algebraic construction is analogous to the paradigmatic Yang's $\eta$-pairing mechanism~\cite{etaPairingYang89,moudgalya2020etapairing,mark2020unified}. We further corroborate this construction via exact diagonalization simulations of the latticized QED$_2^{(q,-q)}$, showing that the analytically predicted scar states remain exact eigenstates under perturbations that thermalize the bulk of the theory's spectrum, and are sharply
distinguished from nearby states by their entanglement and long-range pair correlation.  

The scar construction also admits a direct bosonic interpretation. After charge conjugation of the negatively charged species, the massless theory is mapped to the ordinary two flavor Schwinger model: gauge dynamics gap the total $\mathrm{U}(1)$ charge sector, while the neutral infrared sector is described by  a compact boson~\cite{Coleman:1976uz,Witten:1983ar,Gepner:1984au,Affleck:1985wa,Dempsey:2023gib,Cuomo:2026kmy}. In this representation, 
the scar tower forms a single maximal vector-flavor multiplet and its complete current algebra can be realized entirely within the compact boson. Coherent superpositions of fixed-charge scar states then carry an internal flavor angle whose uniform precession follows 
a Josephson-type relation. 
The same neutral sector supports gapless propagating flavor modes, suggesting that slowly varying deformations of this collective phase can be interpreted as flavor-sound-like excitations.  
Although this construction provides a continuum realization of the scar algebra and its collective dynamics, the maximal lattice multiplet carries a cutoff-scale excitation energy relative to the physical vacuum, and its existence as a finite-energy state in the continuum theory remains a separate question.

\textit{Scar states from group invariant subspaces:} Quantum many-body scars form a subspace $\cal S$ that is dynamically decoupled from the rest of the Hilbert space \cite{Bernien:2017ubn,Turner:2017fxc,Serbyn:2020wys,Moudgalya:2021xlu,Papic2022,Chandran:2022jtd}. This may be a consequence of different Hamiltonians acting on the two subspaces which in turn happens when some terms in the Hamiltonian exactly annihilate the scar subspace. In cases when the scar subspace has certain continuous symmetry $G$ with generators $T_i$ the terms annihilating scar states can be written as $\sum_i O_i T_i$, with $O_i$ arbitrary operators such that $O_i T_i$ is Hermitian~\cite{pakrouski2020GroupInvariantScars}. With the full Hamiltonian given by
\begin{align}
\label{eq:H0PlusOT}
H=H_0+\sum_i O_i T_i\, ,
\end{align}
$H_0$ must respect the symmetry $G$ at least over the subspace $\cal S$ \cite{pakrouski2020GroupInvariantScars}. 

For fermionic lattice models in any dimension many-body scars of the form 
\begin{align}
\label{eq:genTowerWf}
\ket{\phi_n} = \frac{(\sum_i\CP^{\dagger}_i)^n \ket{0_\phi}}{P_L(n)}, \quad 0\le n\le LK
\end{align}
naturally arise for groups $G$ having a sub-group O$(L)$ \cite{paperC1}. Here $2K$ is the number of fermionic flavours per site, $\CP$ is an operator bilinear in fermionic creation and annihilation operators, $\ket{0_\phi}$ is the suitable vacuum state and $P_L(n)$ is a normalization factor.
An alternative to Eq.~\eqref{eq:genTowerWf} basis is given by
\begin{gather}
\ket{z_n} =\frac{(\CP^{\gamma\dagger})^n}{P_L(n)} \ket{z_0},
\label{eq:zntower}
\end{gather}
where $\CP^{\gamma\dagger}$ is a certain transformation of $\CP^{\dagger}$ and the lowest scar state $\ket{z_0}$ is the coherent state
\begin{align}
\label{eq:gsGeneralExp}
|z_0\rangle  
= u^{LK} \prod_j e^{ \frac{v}{u} \CP^\dagger_j}\ket{0_\phi},
\end{align}
with $v$ and $u$ constants depending on $H_0$ only~\cite{paperC1}. In fermionic lattice systems with two flavor species, i.e. $K=1$,
three such group-invariant scar subspaces have been identified \cite{pakrouski2020GroupInvariantScars,Pakrouski:2021jon,2020MarkMotrEtaPairHub,moudgalya2020etapairing} with $\CP^{\dagger}_j$ given by 
\begin{align}
\label{eq:1bandPairingOdagMain}
\eta_j^\dagger  = \chi^\dagger_{j,+} \chi^\dagger_{j,-};  \quad
{\eta '}_j^{\dagger}  =   e^{i\pi j} \chi^\dagger_{j,+} \chi^\dagger_{j,-}; \quad
{\zeta}_j^{\dagger}  = \chi^\dagger_{j,+} \chi_{j,-}\, .
\end{align}
While the three subspaces are distinct with different symmetries $G$ there is a transformation converting $\eta$ to $\eta'$ \cite{Pakrouski:2021jon}, while the Shiba transformation \cite{hubbard1DbookShiba} turns $\eta'$ states to $\zeta$. 

As we show, the $\eta'$ scar family, introduced in Eq.~\eqref{eq:tower_intro} and defined on any bipartite lattice~\cite{etaPairingYang89}, generates a scar subspace in QED$_2^{(q,-q)}$. The generator algebra corresponding to the full symmetry group of these states includes~\cite{Pakrouski:2021jon} nearest-neighbor hopping with real amplitude
\begin{gather}
\label{eq:reHopping}
T'_{jj+1} = \sum_{f=1}^{2}
\left(
\chi^\dagger_{f,j}\chi_{f,j+1}
+\chi^\dagger_{f,j+1}\chi_{f,j}
\right),
\end{gather}
the flavor-dependent hopping (see End Matter [EM])
\begin{gather}
 \label{eq:transformedSOHopping}
{\tilde{T}^{z\prime}_{jj+1}}= -i \sum_{\alpha\beta} \chi^{\dagger}_{\alpha,j} \sigma^{z}_{\alpha\beta} \chi_{\beta,j+1} + \mathrm{h.c.}\, ,
 \end{gather}
and 
\begin{gather}
\label{eq:SiInFerm}
S^A_i=  \frac{1}{2} \sum_{\alpha,\beta} \chi^\dagger_{\alpha,i} \sigma^A_{\alpha \beta} \chi_{\beta,i}\ ,
\end{gather}
where $\sigma^A$ are the Pauli matrices. 
All these terms naturally appear in the Hamiltonians we study in this work. More, using the mapping between scar subspaces in Eq.~\eqref{eq:1bandPairingOdagMain}, one can generate other scar towers in different gauge theories, as we illustrate.

\textit{Singlet scar towers in lattice gauge theories:} The Schwinger model 
with two oppositely charged $q=1$ fermions 
is defined on a staggered  Kogut-Susskind~\cite{Kogut:1979wt,Kogut:1974ag} lattice by the Hamiltonian:
\begin{align}
    H
    ={}&-\frac{1}{2a}\sum_j\left(
        \chi_{+,j}^\dagger U_j\chi_{+,j+1}
        +\chi_{-,j}^\dagger U_j^\dagger\chi_{-,j+1}
        +\mathrm{h.c.}\right)    \nonumber\\
        &+\frac{g^2a}{2}\sum_j E_j^2
        +m_f\sum_j(-1)^j(n_{+,j}+n_{-,j})\nonumber\\
        &+\frac{\Delta}{2}\sum_j(n_{+,j}+n_{-,j}-1),
    \label{eq:Hferm}
\end{align}
where for each link  one has the operator $U_j=e^{i\theta_j}$ and the (dimensionless) electric field operator $E_j=-i\partial_{\theta_j}$ with $[E_j,U_k]=\delta_{jk}U_k$, and $n_{\alpha,j}=\chi_{\alpha,j}^\dagger\chi_{\alpha,j}$. The model is controlled by the homogeneous (vanishing) fermionic masses $m_f$, the gauge coupling $g$ and lattice spacing $a$. A chemical potential-like term is %introduced, 
controlled by the $\Delta$ parameter. Physical states $\ket{\Psi}$ must be chargeless and thus obey Gauss's law $G_j\ket{\Psi}=0$ with $G_j=E_j-E_{j-1}-(n_{+,j}-n_{-,j})$; we use open boundary conditions with fluxless boundaries.

To ascertain that QED$_{2}^{(1,-1)}$ supports scar states, we check that Eq.~\eqref{eq:Hferm} follows group theory arguments laid out for the $\eta'$ family. First, the gauge-covariant hopping operator in Eq.~\eqref{eq:Hferm} is a simple sum of the generators of the full symmetry group of the states $\ket{\eta^{\prime}}$ given in Eqs.~\eqref{eq:reHopping} and \eqref{eq:transformedSOHopping}, see EM. More, integrating out the gauge field from Eq.~\eqref{eq:Hferm} using the Gauss operator, the electric field coincides with the generators $2 \sum_j S^3_j=\sum_j \left(n_{+,j} - n_{-,j}\right)$ in Eq.~\eqref{eq:SiInFerm}. Therefore, the first two terms in Eq.~\eqref{eq:Hferm} are composed of generators only, and can be schematically written as $T+T^2$, in terms of the form in Eq.~\eqref{eq:H0PlusOT}, and exactly annihilate the subspace spanned by $\ket{\eta^{\prime}}$. As a result, this theory has the desired structure to support many-body gauge-invariant scars, as long as $m_f=0$, with the Hamiltonian decomposing exactly into the form in Eq.~\eqref{eq:H0PlusOT} with respect to the symmetry of the $\eta'$ states. However, the $L+1$ generated scar states are energy degenerate, making the structure in their subspace trivial. This degeneracy is lifted by the chemical potential, which is known to preserve the $\ket{\eta^{\prime}}$ subspace
~\cite{pakrouski2020GroupInvariantScars}. As each application of $\eta'^\dagger$ creates
two fermions, this term acts within the tower as $ H_\Delta\ket{\eta'_n}
 =\Delta\left(n-\frac{L}{2}\right)\ket{\eta'_n}$, and
the scar states are thus equally separated in energy by $\Delta$, causing
coherent superpositions within the tower to undergo exact revivals with period $2\pi/|\Delta|$.

When considering a fermionic mass perturbation to the theory, the tower is not conserved as $[\sum_j(-1)^j
\left(n_{+,j}+n_{-,j}\right),\eta^{\prime\dagger}]=2\eta^\dagger$. 
Thus, the mass term 
mixes the protected $\eta^{\prime\dagger}$ operator with
$\eta^\dagger$. For even $L$, the corresponding one-pair states are
orthogonal, i.e. $\bra{\Omega}\eta\,\eta^{\prime\dagger}\ket{\Omega}
    =\sum_{j=1}^{L}(-1)^j=0$, and
the mass deformation generates a component outside the scar subspace. By contrast, symmetry-breaking perturbations can be constructed by sandwiching
arbitrary local operators between the same nearest-neighbor hopping generators, see Eq.~\eqref{eq:reHopping},
that enter the kinetic term. As these generators annihilate every state
$\ket{n^{\eta^\prime}}$ in the scar tower, the resulting perturbations leave the
entire tower unchanged while acting nontrivially on the remaining states. They can therefore render the bulk spectrum quantum chaotic already in small systems without modifying the
scar states or their energies. It is worth pointing out that Hamiltonian~\eqref{eq:Hferm} conserves the total \textit{spin} $J$ and its projection $J_z$ of the \textit{pseudo-spin} SU$(2)$ symmetry generated by ${\eta '}^{\dagger}$, ${\eta '}$ and $[{\eta '},{\eta '}^{\dagger}]$, while the chemical potential does not commute with some of the generators. 
As a result, and without introducing symmetry-breaking perturbations, the highest-$J$ sector comprised of scar states is disconnected from the bulk. This remaining $J$ conservation is not unexpected in the lattice theory, and it can be (also) broken by introducing additional gauge-invariant interactions, see e.g.~\cite{Cherman:2022ecu}, rendering the subspace separation a clean many-body scar phenomenon. In the EM, we discuss a current-current interaction that exactly annihilates and preserves scars but breaks the pseudo-spin symmetry.

\begin{figure}
    \centering
    \includegraphics[width=.9\columnwidth]{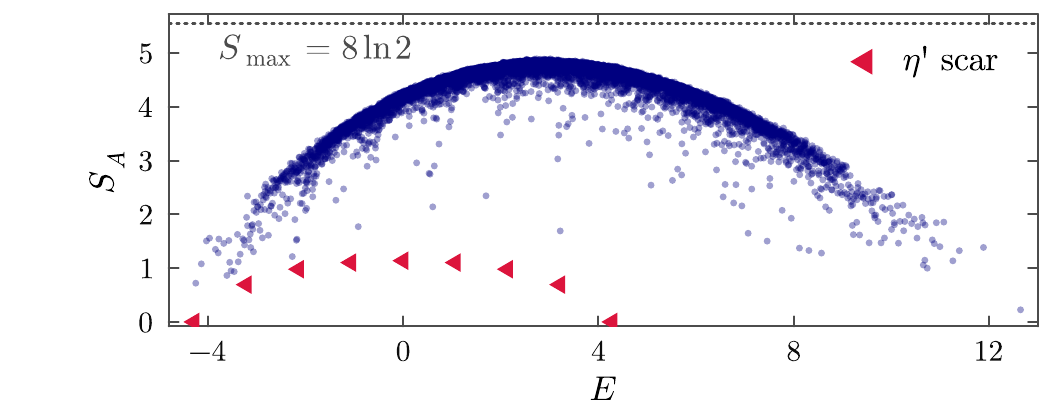}
    \includegraphics[width=.9\columnwidth]{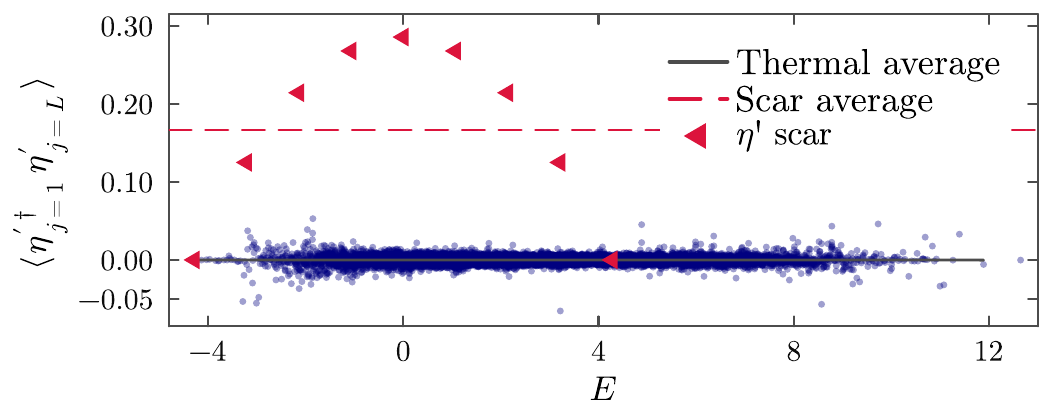}
    \caption{\textbf{Top:} Entanglement entropy for the middle cut. The top horizontal line indicates the maximum possible entanglement in the system. Entanglement of scars is in agreement with analytical expressions~\cite{SciPostPhys.3.6.043}. \textbf{Bottom:} Two-point function $\braket{{\eta^{\prime\dagger}_{j=1}}\eta^{\prime}_{j=L}}$. In the scar states it is $\frac{n(L-n)}{L(L-1)}$~\cite{paperC1}. Dashed red line indicates the average over the scar subspace while the gray line the thermal average over all the eigenstates.}
    \label{fig:prelimNumerics}
\end{figure}

In Fig.~\ref{fig:prelimNumerics} we show supporting numerical evidence for the $\eta'$ scars in QED$_2^{(1,-1)}$. We exactly diagonalize the Hamiltonian in Eq.~\eqref{eq:Hferm}, with an auxiliary symmetry-breaking term 
\begin{gather}
\label{eq:TOT2}
OT =  0.626 \sum_{\braket{i,j}} T_{ij} V_{i} T_{ij},  \\ \nonumber
V_{i} =  r^{(1)}_{i,\alpha\beta}\chi^{\dagger}_{i,\alpha} \chi_{i,\beta} + r_{i,\alpha\beta}^{(2)}\chi^{\dagger}_{i,\alpha} \chi^{\dagger}_{i,\beta} + {\rm h.c.},
\end{gather}
where $r^{(1)}_{i,\alpha\beta}$ and $r^{(2)}_{i,\alpha\beta}$ are real random numbers, uniformly distributed between $-1/2$ and $1/2$, only non-zero for the terms that preserve the total charge, and $T_{ij}$ are the hopping terms. 
This term does not affect the scar subspace, but \textit{thermalizes} the complement subspace, allowing to more clearly distinguish the scar states. The only remaining quantum number is the total charge (i.e. preserving gauge symmetry) and in Fig.~\ref{fig:prelimNumerics} we show the results in the chargeless sector for $L=8$ sites, $a=1$, $\Delta a=1.07$, $m_fa=0$, $ga=0.33$.

The $\ket{n^{\eta^\prime}}$ tower is identified by exact overlap with Eq.~\eqref{eq:tower_intro}. The scar states exhibit anomalously low bipartite entanglement entropy across the middle cut compared with generic eigenstates at similar energies, providing a hallmark of their nonthermal character.
Likewise, the two-point correlations $\braket{{\eta^{\prime\dagger}_{j=1}}\eta^{\prime}_{j=L}}$ representing the off-diagonal long-range order~\cite{yang1962concept} are substantially higher in the scar states compared to generic states at around the same energy, a typical signal of ETH violation.
There are almost no near-zero gaps 
and the level statistics parameter $r=0.5301$ is very close to the GOE value of 0.5359 indicating that the bulk of the spectrum is fully ergodic without remaining symmetries and in presence of the auxiliary $TOT$ term in Eq.~\eqref{eq:TOT2}. We thus observe \textit{genuine} scars, i.e. the observed states Hilbert space fraction vanishes exponentially with $L$, while their pair correlations remain
athermal.

It is natural to wonder how this construction can be generalized to other two dimensional gauge theories. It is evident that such scar towers can not be constructed in one flavor QED, as it is not possible to build any of the subspace generators in Eq.~\eqref{eq:1bandPairingOdagMain}. Equivalently, for odd number of flavors, the scar states can exist as long as one selects an even flavor sub-sector, and thus it is sufficient to discuss the case with two flavors, albeit more complex subspaces can be constructed for larger $K$. The direct generalization to non-Abelian models is possible, and we leave it for future work.

Consider
two fermions with charges $q_1$ and $q_2$, such that their commutations with the local charge operator read
  $  [\hat Q,\chi_{1,j}^\dagger\chi_{2,j}^{[\dagger]}]
    =(q_1-[+] \,q_2)\chi_{1,j}^\dagger\chi_{2,j}^{[\dagger]}$. For $q_1/q_2=-1$, the particle--particle operators $\eta_j^\dagger$ and ${\eta'_j}^\dagger$ are gauge neutral, whereas the flavor-flip operator $\zeta_j^\dagger$ is charged, and thus can not be a generator of scar subspace. The real gauge-covariant hopping of QED$_2^{(q,-q)}$ selects the staggered $\eta'$ tower constructed above. The unstaggered $\eta$ tower would instead require the corresponding hopping with a purely imaginary amplitude.
For $q_1/q_2=1$, the situation is exactly reversed: pair creation carries nonzero gauge charge, while $\zeta^\dagger
=\sum_j\chi_{1,j}^\dagger\chi_{2,j}$ is a local gauge-neutral flavor rotation. Thus in the equally charged model only the $\zeta$ scar tower exists, while the $\eta$ and $\eta'$ generators do not generate a protected subspace. Nonetheless, these operators are connected by charge conjugation of the negatively charged species (Shiba transformation),
$\left\{c_{1,j}=\chi_{+,j}, c_{2,j}=(-1)^j\chi_{-,j}^\dagger\right\}$,
under which the original $\eta'$ operator transforms as
\begin{equation}
\eta^{\prime\dagger}
=\sum_j(-1)^j\chi_{+,j}^\dagger\chi_{-,j}^\dagger
\longrightarrow
\sum_jc_{1,j}^\dagger c_{2,j}
=\zeta^\dagger .
\label{eq:eta_zeta_mapping}
\end{equation}
As a result, the $\eta'$ tower in QED$_2^{(q,-q)}$ corresponds to the $\zeta$ tower in QED$_2^{(q,q)}$, trading the different flavors by a neutral flavor polarized operator. The $\eta$ tower is never realized.

\begin{figure}
    \centering
    \includegraphics[width=1\columnwidth]{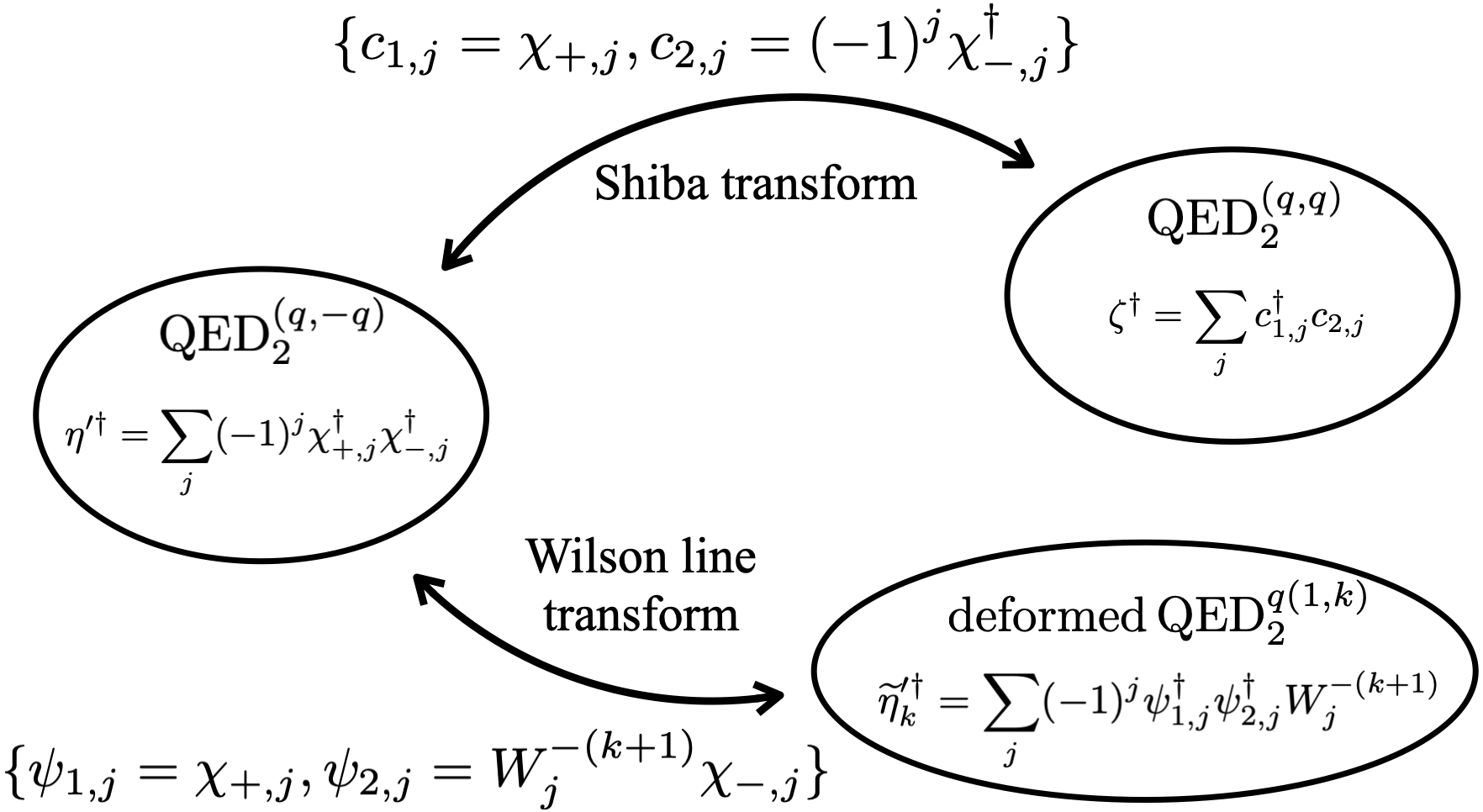}
    \caption{Mappings of the $\eta'$ scar tower under Shiba and Wilson line transformations.}
    \label{fig:cartoon}
\end{figure}

Finally for $(q_1,q_2)=q(1,k)$, with $k\neq \pm 1,0$ an integer number, we consider the field redefinitions of the fermions 
through the string operator $W_j=\prod_{\ell<j}U_\ell$:
\begin{equation}
\psi_{1,j}=\chi_{+,j},
\qquad
\psi_{2,j}=W_j^{-(k+1)}\chi_{-,j}\, .
\label{eq:generic_charge_field_map}
\end{equation}
The Wilson string attaches $(k+1)$ units of electric flux to the second fermion, thereby changing its charge from $-q$ to $kq$. In particular, using $W_{j+1}=W_jU_j$, its hopping term transforms as $\chi_{-,j}^\dagger U_j^\dagger\chi_{-,j+1}
\longrightarrow
\psi_{2,j}^\dagger U_j^k\psi_{2,j+1}$, which is the gauge-covariant hopping for a fermion of charge $kq$. The local neutral pair of the opposite-charge theory is simultaneously mapped to a string-dressed operator, $\chi_{+,j}^\dagger\chi_{-,j}^\dagger
\longrightarrow
\psi_{1,j}^\dagger\psi_{2,j}^\dagger
W_j^{-(k+1)}$. Thus, although the image of the original scar subspace remains well defined under the full transformation, its generator is nonlocal. More importantly, the transformation acts nontrivially on the electric energy, which becomes
$\frac{g^2a}{2}\sum_\ell
\left[
E_\ell+(k+1)\sum_{m>\ell}n_{2,m}
\right]^2$ 
with $n_{2,m}=\psi_{2,m}^\dagger\psi_{2,m}$, see Eq.~\eqref{eq:Hferm}. 
Consequently, the transformed Hamiltonian is a \textit{deformed} QED$_{2}^{q(1,k)}$ gauge theory containing additional density-dependent couplings to the electric flux. The original fluxless sector maps to $\mathcal E_0=-(k+1)N_2$ and $\mathcal E_L=0$,
where $N_2=\sum_j n_{2,j}$.  Indeed, for an $n$-pair scar
$N_1=N_2=n$ and $\mathcal E_L-\mathcal E_0=(k+1)n$, so the
target theory is understood with a compensating boundary charge. The relation among these theories and their associated scar operators is summarized in Fig.~\ref{fig:cartoon}.

\textit{Many-body interpretation of the scars:}
The Shiba transformation in Eq.~\eqref{eq:eta_zeta_mapping} gives the scar tower a direct many-body interpretation. In the transformed variables, the spectrum-generating operator is the gauge-neutral flavor-raising operator $\zeta^\dagger$ introduced above. Its continuum counterpart is the raising charge
of the physical vector-flavor symmetry, $Q^+_{V,F}=\int dx\,J^{+,0}_{V,F}(x)$. Here
$J^{+,\mu}_{V,F}=J^{1,\mu}_{V,F}+iJ^{2,\mu}_{V,F}$, with
$J^{a,\mu}_{V,F}=\bar{\Psi}\gamma^\mu\sigma^a\Psi/2$
and $\Psi=(\psi_1,\psi_2)^T$ the transformed flavor doublet.
Thus, after the Shiba transformation the scar tower is a physical
vector-flavor multiplet.

In the massless continuum theory, bosonization separates the dynamics
into a massive flavor-singlet scalar and a massless neutral flavor scalar~\cite{Coleman:1976uz},
\begin{align}
\mathcal{L}_{\mathrm{bos}}
&=
\frac{1}{2}
(\partial\phi_\Sigma)^2
-
\frac{1}{2}m_\Sigma^2\phi_\Sigma^2
+
\frac{1}{2}
(\partial\phi_F)^2\,,
\end{align}
where $m_\Sigma^2=\frac{2g^2}{\pi}$. Writing the bosons associated with the two transformed flavors as
$\phi_1$ and $\phi_2$, the fields
$\phi_\Sigma=(\phi_1+\phi_2)/\sqrt{2}$ and
$\phi_F=(\phi_1-\phi_2)/\sqrt{2}$ describe, respectively, the
flavor-singlet charge sector and the neutral flavor sector.
The gauge field couples to the total $\mathrm{U}(1)$ charge and generates
the Schwinger mass, while the nonsinglet flavor sector is gauge neutral
and anomaly free and therefore remains gapless. Its infrared description
is the $\mathrm{SU}(2)_1$ Wess--Zumino--Witten theory, equivalently a compact
boson~\cite{Coleman:1976uz,Witten:1983ar,Gepner:1984au,Affleck:1985wa}.
Defining similarly
$J^{a,\mu}_{A,F}=\bar{\Psi}\gamma^\mu\gamma^5\sigma^a\Psi/2$,
the Cartan components of the physical vector and axial flavor currents are
\begin{align}
J^{3,\mu}_{V,F}
&=
\frac{1}{\sqrt{2\pi}}
\epsilon^{\mu\nu}\partial_\nu\phi_F,
&
J^{3,\mu}_{A,F}
&=
\frac{1}{\sqrt{2\pi}}
\partial^\mu\phi_F.
\end{align}
The transverse components are the corresponding $\mathrm{SU}(2)_1$
vertex operators, built from the same compact field $\phi_F$ and its
dual~\cite{Halpern:1975jc}. The complete physical flavor-current algebra containing the scar
generator can therefore be realized entirely within the compact neutral
boson.

This identification also makes the structure of the exact tower
transparent. The Shiba image of the reference state is the lowest-weight
state $\ket{J,-J}$ of the maximal vector-flavor representation, with
$J=L/2$, and
\begin{align}
\ket{n^{\eta^\prime}}
\propto
\left(Q^+_{V,F}\right)^n
\ket{J,-J}.
\end{align}
The $n^{\rm th}$ state carries
$Q^3_{V,F}=n-L/2$, while all states in the tower have the same
vector-flavor Casimir
$\boldsymbol{Q}_{V,F}^{\,2}=J(J+1)$.
Thus, changing $n$ changes only the Cartan projection within the maximal
representation. In the compact-boson language, $Q^3_{V,F}$ measures the
winding of $\phi_F$. The winding alone, however, does not specify the
scar state, since the fixed maximal Casimir also carries transverse
flavor information. In particular, the central scar has
$Q^3_{V,F}=0$ but cannot be identified with the neutral-boson vacuum,
which belongs to a different flavor representation.

The $\eta'$ construction can also be expressed directly in bosonic
variables. The inverse Shiba transformation reflects the second flavor
boson and its dual. Thus, in the original opposite-charge theory, the
neutral compact field is the normalized sum of the two flavor bosons,
while their difference belongs to the massive charge sector. The same
vertex-current construction then realizes the pseudospin
$\mathrm{SU}(2)$ entirely within the neutral boson, with the $\eta'$
tower corresponding to its $J=L/2$ multiplet. The raising operation
leaves the massive charge sector unchanged.

The coherent state introduced in Eq.~\eqref{eq:gsGeneralExp} has an
equally direct interpretation as an $\mathrm{SU}(2)$ flavor-coherent
state. Writing $z=v/u=\rho e^{-i\vartheta}$, it takes the form
\begin{align}
\ket{z;J}
=
\frac{e^{zQ^+_{V,F}}}
{\left(1+|z|^2\right)^J}
\ket{J,-J}.
\end{align}
Its transverse flavor polarization is
\begin{align}
\left\langle Q^+_{V,F}\right\rangle
=
\frac{2J\rho}{1+\rho^2}
e^{i\vartheta}.
\end{align}
The transverse polarization is therefore extensive in the system size,
and $\vartheta$ describes a collective internal flavor orientation.
By contrast, every fixed-$n$ state has
$\langle Q^+_{V,F}\rangle=0$ and no definite transverse orientation:
the collective angle becomes meaningful only after coherently
superposing different vector-charge sectors.

The equally spaced scar spectrum gives this collective angle an exact
dynamics. The chemical-potential-like term $H_\Delta$ introduced above
takes a particularly simple form in the transformed variables, $H_\Delta=\Delta Q^3_{V,F}$. Then, $[H,Q^\pm_{V,F}]=\pm\Delta Q^\pm_{V,F}$, since the remaining massless
Hamiltonian is vector-flavor symmetric. Using the scar energies, the coherent state consequently evolves as $e^{-iHt}\ket{z;J}=e^{i\Delta Jt}\ket{ze^{-i\Delta t};J}$. The overall phase is immaterial, while the collective flavor angle
satisfies
\begin{align}
\dot{\vartheta}
=
\Delta.
\end{align}
This exact phase--charge relation provides a Josephson-type many-body
interpretation of the scar oscillations, analogous to the phase--chemical-potential relation for the collective chiral mode of massless fermions in two dimensions~\cite{Mottola:2019nui}. The parameter $\Delta$, which
fixes the uniform level spacing of the tower, acts as the
chemical-potential-like bias conjugate to the vector-flavor charge,
while the coherent scar undergoes uniform precession of its transverse
flavor polarization. The bosonic Shiba reflection maps the flavor-current
algebra of the $\zeta$ tower to the pseudospin-current algebra of the
$\eta'$ tower, both realized by the neutral compact boson. It thereby
identifies the angle of the transverse flavor polarization with the
phase of the coherent pair amplitude, giving the same precession law
in both descriptions.

The unstaggered $\eta$ operator instead maps under Shiba to a staggered flavor bilinear whose bosonized expression involves both $\phi_F$ and $\phi_\Sigma$. It is not a raising charge of the neutral current algebra and does not generate a protected tower for the hopping considered here. However, a sublattice phase rotation maps the protected $\eta'$ tower to the unstaggered $\eta$ tower of the correspondingly transformed Hamiltonian, whose hopping prefactors are purely imaginary. Bosonization adapted to this transformed kinetic term again realizes the protected generator entirely within the neutral current algebra. Consequently, coherent superpositions of the transformed $\eta$ tower exhibit the same Josephson-type many-body precession.

The same compact neutral sector supports gapless propagating flavor
excitations with linear dispersion. This suggests that slowly varying
deformations of the coherent flavor orientation may be related to
flavor-wave, or flavor-sound-like, modes of the massless theory.
Establishing this relation requires deriving the long-wavelength
effective dynamics directly around the coherent scar state, since the
latter also carries the nontrivial maximal-representation information
described above. In addition, and as shown above, the mass term
mixes $\eta^{\prime\dagger}$ with the orthogonal $\eta^\dagger$ channel
and therefore breaks the exact spectrum-generating algebra. Weak mass
deformations consequently provide a natural setting in which to study
the crossover from exact Josephson-like scar oscillations to long-lived
approximate dynamics. 

Finally, the continuum realization of the flavor algebra and its
collective precession should be distinguished from a finite-energy
continuum limit of the particular maximal lattice tower. At fixed
physical length $\ell=La$, the size of the maximal representation grows
with the ultraviolet cutoff. Already in the free theory with $\Delta=m_f=0$, for an open chain even $L$,
$E_{\mathrm{scar}}-E_{\mathrm{vac}}=
2\ell/(\pi a^2)+O(1/a)$, so the tower remains separated from the
physical vacuum by a cutoff-scale energy. What has an immediate
continuum realization is therefore the physical flavor-current algebra,
the identification of the scar generator with the vector-flavor raising
charge, and the associated Josephson-like collective dynamics. Whether
the same spectrum-generating structure admits a finite-energy
realization above the physical vacuum is an open
question.

\textit{Conclusion:} We have constructed an exact tower of gauge-invariant many-body scar states in two-dimensional gauge theories with two massless fermion flavors. For oppositely charged fermions, the tower is generated algebraically by a gauge-neutral staggered $\eta^\prime$-pairing operator, as confirmed by exact diagonalization tests of the anomalously low entanglement and off-diagonal long-range pair correlations. Under a Shiba transformation, the tower maps onto a maximal vector-flavor multiplet of the equal-charge theory, providing an interpretation in terms of the compact neutral boson. Coherent superpositions of the scar states consequently exhibit Josephson-like flavor precession. We have further shown that finite fermion masses mix the protected pair with an orthogonal channel and destroy the exact tower, while a broad class of symmetry-breaking interactions can thermalize the bulk without affecting the scars. The discussion here can be generalized to multiflavor models with non-Abelian gauge symmetries, and we expect part of the presented arguments to also apply in higher dimensions. These results provide, to our knowledge, the first analytical construction of an invariant scar subspace in a gauge theory with a nontrivial continuum interpretation.

\textit{Acknowledgment:} KP is supported by the Swiss National Science Foundation, Spark grant CRSK-$2\_237767$. The work of AVS is supported by State Agency for Research of the Spanish Ministry of Science and Innovation through grant PID2025-174203NB-I00 and by the Basque Government through grant IT1977-26. AVS would also like to acknowledge support from Ikerbasque, Basque Foundation for Science. We are also grateful for the organization and participation in the CERN workshop "Chaos, Scars, Thermalization", where this work was initiated.

\section{End Matter}

\subsection{Details of numerical computation}
After integrating out the gauge field, the Hamiltonian used for the numerical study in the main text is written as
\begin{align}
H
={}&-\frac{1}{2a}
\sum_{j=1}^{L-1} T'_{jj+1}+\sum_{j=1}^{L}\sum_{f=1}^{2}
m_f(-1)^j \, n_{f,j}
\nn
&\hspace{-1 cm}+\frac{g^2a}{2}
\sum_{\ell=1}^{L-1}
\left[
\sum_{j=1}^{\ell}
\left(2S^3_j\right)
\right]^2+\frac{\Delta}{2}
\sum_{j=1}^{L}
\left(n_{1,j}+n_{2,j}-1\right)\, .
\label{eq:Hfermion}
\end{align}
This form is obtained after using Gauss's law to derive the relation
\begin{equation}
E_\ell=\sum_{j=1}^{\ell}\left(n_{1,j}-n_{2,j}\right)
=\sum_{j=1}^{\ell}2S_j^3\, ,
\end{equation}
and the link variables in the hopping term are removed by a gauge transformation, resulting in the real hopping \(T'_{jj+1}\). We further impose that physical states are in the chargeless sector of the theory.

\subsection{The transformation from $\ket{n^\eta}$ to $\ket{n^{\eta^\prime}}$}

The mapping \(R\), which relates \(\ket{n^\eta}\) to $\ket{n^{\eta^\prime}}$~\cite{Pakrouski:2021jon}, introduces the alternating sign structure of the latter states and is defined only on a bipartite lattice with sublattices \(A\) and \(B\). It acts trivially on sublattice \(A\), while on sublattice \(B\) it acts as
\begin{gather}
\label{eq:transformSimpleBInverse}
\tilde{\chi}_{j,\sigma} \rightarrow -i \chi_{j,\sigma} \, , \quad 
\tilde{\chi}_{j,\sigma}^\dagger \rightarrow  i \chi^\dagger_{j,\sigma} \, .
\end{gather}
Under this transformation, the imaginary-amplitude nearest-neighbor hopping, the operator \(\eta^+\), and the states \(\ket{n^\eta}\), expressed in terms of the \(\tilde{\chi}\) fields, are mapped respectively to real-amplitude hopping, the operator \(\eta^{+\prime}\), and the states \(\ket{n^{\eta^\prime}}\), expressed in terms of the \(\chi\) fields. Consequently, real-amplitude hopping annihilates \(\ket{n^{\eta^\prime}}\), whereas imaginary-amplitude hopping annihilates \(\ket{n^\eta}\). As a result, the $\eta$ tower is not realized in \textit{standard} lattice gauge theories.

\subsection{The gauge-covariant hopping annihilates $\ket{n^{\eta^\prime}}$}

 The gauge-covariant hopping in Eq.~\eqref{eq:Hferm} can be written as  
\begin{align}
\label{eq:HkinTheta}
H_{kin} &= -t \sum_{j,\sigma} \cos(\theta_j) \chi^{\dagger}_{j,\sigma} \chi_{j+1,\sigma}   \\ \nonumber
&-i t \sum_{j,\alpha,\beta} \sin(\theta_j) \chi^{\dagger}_{j,\alpha} \sigma^z_{\alpha\beta} \chi_{j+1,\beta} + \mathrm{h.c.}\, ,
 \end{align}
where we used $U_j = e^{i\theta_j}$. In the first line, the real hopping is known to annihilate $\ket{n^{\eta^\prime}}$ \cite{Pakrouski:2021jon}. The $\ket{n^\eta}$ states are annihilated by the \textit{spin-orbit} coupled hopping $\tilde{T}^A_{ij}=\sum_{\alpha\beta}\chi^\dagger_{i\alpha}\sigma^A_{\alpha\beta}\chi_{j\beta}+\mathrm{h.c.}$, i.e. Eq.~(13) in~\cite{Pakrouski:2021jon}.
Under the inverse transformation in Eq.~\eqref{eq:transformSimpleBInverse} the third component of this hopping $\tilde{T}^z_{jj+1}$ (see Eq.~\eqref{eq:transformedSOHopping}) coincides with the imaginary part of the gauge-covariant hopping in Eq.~\eqref{eq:HkinTheta}. Since $\tilde{T}^z_{ij}$ annihilates $\ket{n^\eta}$ states, its transformed version (and therefore the gauge-covariant hopping altogether) annihilates the $\ket{n^{\eta^\prime}}$ states.

\subsection{Example operator that breaks the residual pseudo-spin symmetry}

Motivated by~\cite{Cherman:2022ecu} (see also discussion in~\cite{Barata:2023jgd}), we consider the four-fermion current--current interaction. Introducing the flavor doublet $\Psi = (\psi_{+},\psi_{-})^T$, we define
\begin{align}
    J_1^\mu
    &=
    \bar{\Psi}\gamma^\mu \mathbf{1}\Psi
    =
    j_+^\mu+j_-^\mu,
    \nn 
    J_3^\mu
    &=
    \bar{\Psi}\gamma^\mu \sigma^3\Psi
    =
    j_+^\mu-j_-^\mu,
\end{align}
where $j_\pm^\mu
    =
    \bar{\psi}_\pm\gamma^\mu\psi_\pm$. We then add the Thirring-like interaction
\begin{align}
    \delta H
    =
    \lambda
    \int dx\,
    J_1^\mu(x)J_{3,\mu}(x).
\end{align}
Since
\begin{align}
    J_1^\mu J_{3,\mu}
    =
    j_+^\mu j_{+,\mu}
    -
    j_-^\mu j_{-,\mu},
\end{align}
this breaks the residual pseudospin symmetry.
For the staggered lattice formulation, we define
\begin{align}
    Q_j
    &=
    n_{1,j}-n_{2,j}
    =
    2S_j^3,
    &
    F_j
    &=
    n_{1,j}+n_{2,j}-1 .
\end{align}
The standard Kogut--Susskind discretization of the current--current interaction is then
\begin{align}
    \delta H_{\mathrm{lat}}
    =
    \frac{4\lambda}{a}
    &\sum_{j=1}^{L-1}
    \bigg[
        \left(n_{1,j}-\frac12\right)
        \left(n_{1,j+1}-\frac12\right)
       \nn& -
        \left(n_{2,j}-\frac12\right)
        \left(n_{2,j+1}-\frac12\right)
    \bigg].
\end{align}
Equivalently,
\begin{align}
    \delta H_{\mathrm{lat}}
    =
    \frac{2\lambda}{a}
    \sum_{j=1}^{L-1}
    \left(
        F_jQ_{j+1}
        +
        Q_jF_{j+1}
    \right).
\end{align}

For every state \(\lvert n\rangle\) in the scar tower, $Q_j\lvert n\rangle=0$ for all $j$, because each site is either empty or occupied by a neutral pair.  It
therefore follows that
\begin{align}
    F_jQ_{j+1}\lvert n\rangle
    =
    Q_jF_{j+1}\lvert n\rangle
    =
    0,
\end{align}
and hence $ \delta H_{\mathrm{lat}}\lvert n\rangle=0$. At the same time, the local operators \(F_j=2\eta_j^z\) are not global pseudo-spin scalars. Consequently, for generic \(\lambda\), $\left[ \delta H_{\mathrm{lat}},   \boldsymbol{\eta}^{\,2} \right] \neq 0$. Thus the interaction breaks the residual pseudo-spin symmetry in the bulk Hilbert space while preserving the scar tower exactly. We have numerically checked that for $L=8$ and in the charge-less, half-filled sector this, without the $OT$ term in Eq.\eqref{eq:TOT2}, reduces the original SU(2) but possibly retains a smaller symmetry associated to lattice realization of the theory. All the remaining symmetries are completely eliminated for $L=8$ when using 
\begin{align}
    \delta \tilde H_{\mathrm{lat}}
    =
    \frac{2\lambda}{a}
    \sum_{j=1}^{L-1}
    \left(
        F_jQ_{j+1}
        +
        Q_jF_{j+1}
    \right)
    (-1)^j\bigl(n_{1,j}+n_{2,j}\bigr)\, ,
\end{align}
and could be achieved by introducing other higher mass dimension operators or suitable $OT$ terms.

\bibliographystyle{apsrev4-2}
\bibliography{references.bib}

\end{document}